\pdfoutput=1
\documentclass[aps,prd,twocolumn,epsf,superscriptaddress,nofootinbib]{revtex4}
\usepackage{amsfonts}
\usepackage{amsmath}
\usepackage{amssymb,amsbsy}
\usepackage{bm}
\usepackage{makeidx}
\usepackage{latexsym}
\usepackage{graphicx}
\usepackage{epsfig}
\usepackage{subfigure}
\usepackage{dsfont}
\usepackage{mathcomp}
\usepackage{color}
\usepackage{mathrsfs}
\usepackage{epsf}
\usepackage{pstricks}
\newcommand{\beq}{\begin{eqnarray}}
\newcommand{\eeq}{\end{eqnarray}}
\newcommand{\be}{\begin{equation}}
\newcommand{\ee}{\end{equation}}

\newcommand{\gapp}{\mathrel{\raise.3ex\hbox{$>$}\mkern-14mu
              \lower0.6ex\hbox{$\sim$}}}
\newcommand{\lapp}{\mathrel{\raise.3ex\hbox{$<$}\mkern-14mu
              \lower0.6ex\hbox{$\sim$}}}

\makeatletter

\begin{document}
\title{Unifying inflation and dark matter with the Peccei-Quinn field:\\ observable axions and observable tensors}
\author{Malcolm Fairbairn\footnote{malcolm.fairbairn@kcl.ac.uk}}
\affiliation{Physics, Kings College London, Strand, London WC2R 2LS, UK}
\author{Robert Hogan\footnote{robert.hogan@kcl.ac.uk}}
\affiliation{Physics, Kings College London, Strand, London WC2R 2LS, UK}
\author{David J. E. Marsh \footnote{dmarsh@perimeterinstitute.ca}}
\affiliation{Perimeter Institute, 31 Caroline St N, Waterloo, ON, N2L 6B9, Canada}

\begin{abstract}
\noindent
A model of high scale inflation is presented where the radial part of the Peccei-Quinn (PQ) field with a non-minimal coupling to gravity plays the role of the inflaton, and the QCD axion is the dark matter. A quantum fluctuation of $\mathcal{O}(H/2\pi)$ in the axion field will result in a smaller angular fluctuation if the PQ field is sitting at a larger radius during inflation than in the vacuum. This changes the effective axion decay constant, $f_a$, during inflation and dramatically reduces the production of isocurvature modes. This mechanism opens up a new window in parameter space where an axion decay constant in the range $10^{12}\text{ GeV}\lesssim f_a\lesssim 10^{15}\text{ GeV}$ is compatible with observably large $r$. The exact range allowed for $f_a$ depends on the efficiency of reheating. This model also predicts a minimum possible value of $r=10^{-3}$. The new window can be explored by a measurement of $r$ possible with \textsc{Spider} and the proposed CASPEr experiment search for high $f_a$ axions.
\end{abstract}

\maketitle
\section{Introduction}
\noindent In the last few years there has been a lot of excitement among inflationary cosmologists. With the release of the {\em Planck} data \cite{PlanckInflation} and the recent controversy surrounding the BICEP2 data \cite{BICEP2} we have been faced with the serious possibility of model discrimination, and ensuing debates about what this means for inflationary theory \cite{Guth:2013sya,Burgess:2013sla,Ijjas:2014nta,Burgess:2014lza,Dodelson:2014exa}. Of the many inflationary parameters to be constrained perhaps the most crucial one for model builders is the the tensor-to-scalar ratio,
\begin{equation}
r_k=\frac{A_t (k)}{A_s (k)},
\end{equation}
where $A_t (k)$, and $A_s (k)$ are, respectively, the amplitude of tensor and (adiabatic) scalar perturbations at scale $k$, with 
\begin{equation}
A_s = \frac{1}{2 \epsilon} \left( \frac{H}{2 \pi M_{pl}}\right)^2 \hfill ,\quad \hfill A_t = 8 \left( \frac{H}{2 \pi M_{pl}}\right)^2,
\end{equation}
where $H$ is the Hubble scale during inflation, $\epsilon=-\dot{H}/H^2$ is the first slow-roll parameter and $M_{pl}=2.435 \times 10^{18} $ GeV is the reduced Planck mass. We can therefore see that  $r=16 \epsilon$ so combining a constraint on $r$ with the measurement of $A_s= (2.196\pm 0.060) \times 10^{-9}$ \cite{PlanckInflation} we may constrain $H$. A constraint on $H$ is of utmost importance because it can be used to rule out different models of inflation and particle cosmology. In particular, it can have profound consequences for the cosmology of axions \cite{Fox:2004kb,Hertzberg0807,Marsh1403,Visinelli1403}.
\begin{figure}[tb]
\vspace{-0.2em}\includegraphics[width=\columnwidth]{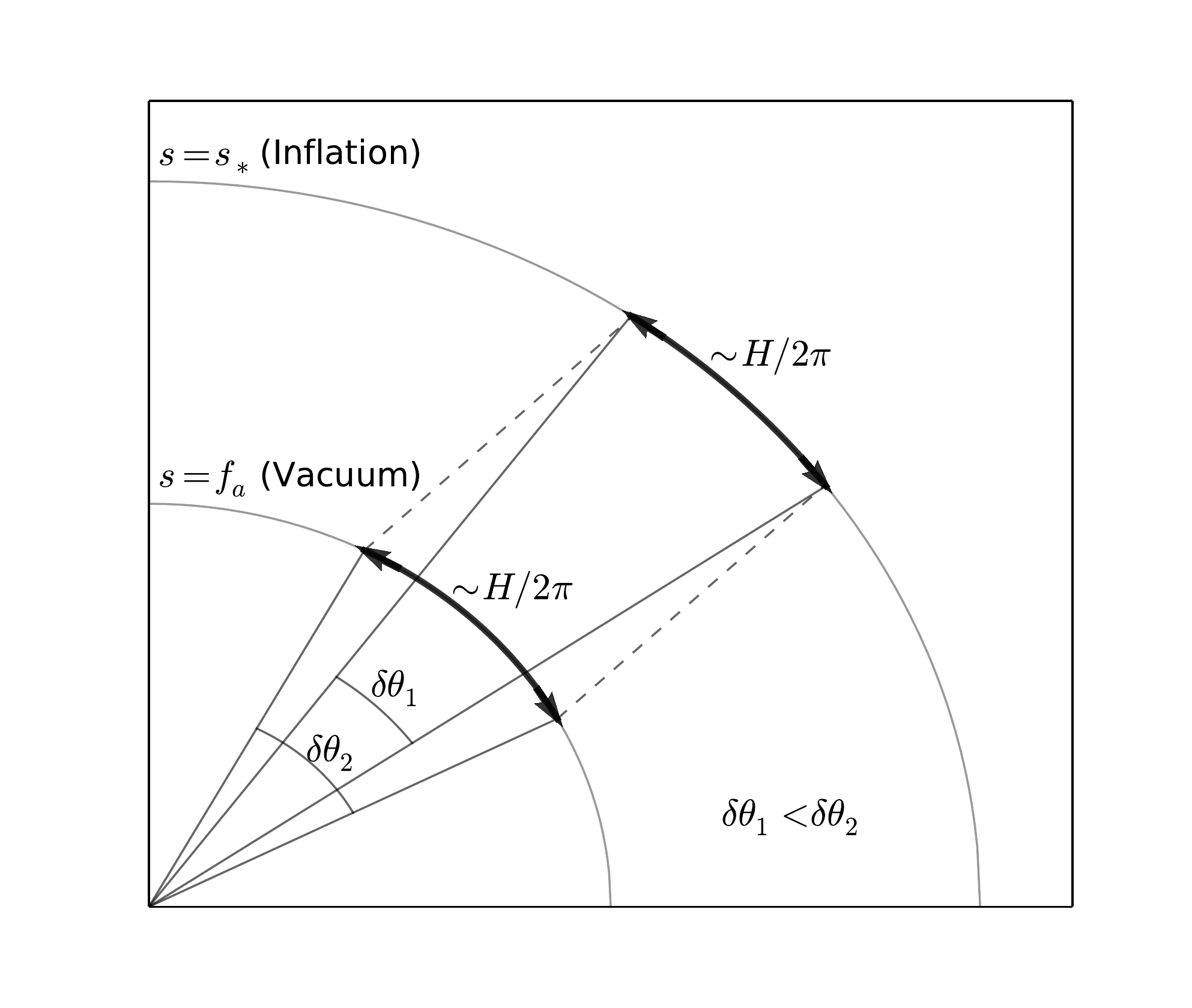}
\vspace{-2.5em}\caption{{\it Schematic of our mechanism. Isocurvature fluctuations in the axion field, $\delta \theta$, are reduced if the radial field,  $s$, lies at higher values during inflation, $s_\star$, compared to the low energy minimum, $f_a$.}}\label{fig:schematic} \vspace{-1.7em}
\end{figure}

The axion \cite{Weinberg:1977ma,Wilczek:1977pj} has long been considered a promising alternative to thermal WIMP dark matter (DM). Axion relics can be produced via the misalignment mechanism where axion particles are produced through oscillations about the symmetry breaking potential minimum \cite{Preskill83,Abbott83,Dine83,Turner83,Turner832,Turner85,Kim87,Berezhiani92}. However, if the energy scale of inflation is high this generates unacceptably large \cite{PlanckInflation} axion isocurvature perturbations if the axion Peccei-Quinn \cite{PecceiQuinn97} (PQ) scale, $f_a$, is larger than $\mathcal{O}(H)$. It appears that only situations where the PQ symmetry is restored after inflation are compatible with an observable $r$. Large $f_a$ models (which are common in top-down approaches such as string theory \cite{Svrcek:2006yi}) would appear to be ruled out if $r$ is observed.

In this paper we show that if the PQ field itself plays the role of the inflaton then the problem of isocurvature modes can be dramatically reduced allowing for high $f_a$ axion DM to be compatible with high scale inflation. 

Fig.~\ref{fig:schematic} shows schematically that if the radial part of the PQ field, $s$, lies at values larger than $f_a$ during inflation then the isocurvature fluctuations of the axion field will be reduced in amplitude. Isocurvature amplitude is proportional to the ratio $\delta\theta/\theta$, where $\theta$ is the axion (angular) direction of the PQ field. The DM abundance fixes $\theta$. Inflation fixes the dimensionful field displacement at $H/2\pi$, however this subtends a smaller angle $\delta \theta$ if it is fixed at large rather than small radius (see Fig.~\ref{fig:schematic}).

In Ref.~\cite{BICEP2} a purported measurement of $r=0.2$ was reported, implying $H\sim 10^{14}\text{ GeV}$. However, recent analyses have called the primordial origin of this signal into question \cite{Flauger1405,Mortonson1405}.  Polarised foreground maps recently released by {\em Planck} \cite{Adam:2014oea} seem to confirm these suspicions and point to the BICEP2 signal being largely due to polarised galactic dust emission. However, there is still room for $0.01\lesssim r\lesssim 0.1$ to be observable and consistent with current constraints. Such a detection could be made, for example, by \textsc{Spider} \cite{2014arXiv1407.2906R}, and the consequences for inflationary cosmology and the axion would still be just as profound \cite{Fox:2004kb}\footnote{An ultimate, cosmic variance limited, measurement of $r$ using 21cm lensing could in principle reach $r\sim 10^{-9}$ \cite{Sigurdson:2005cp,Book:2011dz}. As we will see, even this would provide a constraint on axion physics.}. The model we present here is therefore relevant to axion DM if any measurement of primordial $B$-modes occurs. We do not fix a value of $r$, and consider our model across the entire observable window.

The rest of the paper is organised as follows: in Section~\ref{sec:inflation} we introduce the model of inflation, in Section~\ref{sec:axion_constraints} we discuss the constraints from the axion sector, and in Section~\ref{sec:conclusions} we present our results and conclusions.

\section{Inflation with the Radial PQ Field}
\label{sec:inflation}

The PQ symmetry was first introduced \cite{PecceiQuinn97} to solve the Stong-\emph{CP} problem of the Standard Model. The origin of this problem is the presence of the \emph{CP}-violating topological $\theta$-term,
\begin{equation}
S_{\theta}=\frac{\theta}{32 \pi^2} \int d^4 x \ \text{Tr}\    G^{\mu \nu}\widetilde{G}_{\mu \nu}\, .
\end{equation}
This term generates a electric dipole moment for the neutron which is very tightly constrained ($d_n < 2.9 \times 10^{-26} e\ \text{cm}$ \cite{Baker97}) implying that $\theta$ must be tuned to be very small ($\lesssim 10^{-10}$). The solution provided by the  PQ mechanism is to identify $\theta$ with a pseudo Nambu-Goldstone boson (the axion) of a broken $U(1)$ symmetry. Non-perturbative QCD effects at $T<\Lambda_{\rm QCD}$ generate the axion potential \cite{Gross:1980br} 
\begin{equation}
V(a) \simeq m_a^2 f_a^2 (1- \cos \theta)\, ,
\end{equation} 
where $m_a$ is the axion mass, $f_a$ is the axion decay constant, and $\theta=a/f_a$. This potential causes the $\theta$-term to dynamically relax to zero.\footnote{The particular $(1-\cos\theta)$ form of the potential can vary, but its $CP$-conserving properties are guaranteed \cite{Vafa:1984xg}.}

The PQ field, $S$, is a complex field charged under a global $U(1)$ symmetry broken at scale $f_a$. The axion, $a$, is the angular part of this field. The radial part, $s$, is minimised at $f_a$. Our mechanism for reducing isocurvature perturbations works by taking $s\gg f_a$ during inflation. One mechanism by which this can be achieved is to take $s$ to be the inflaton. 

The usual potential for the PQ field is given by:
\begin{equation}
V=\lambda \left( S^\dagger S -\frac{f_a^2}{2} \right)^2 =\frac{1}{4}\lambda \left( s^2 -f_a^2 \right)^2\, ,
\label{eqn:PQ_potential}
\end{equation}
At large $s$ this takes the form of a $\lambda\phi^4$ single-field inflation model.\footnote{The axion direction is massless during inflation in QCD and we consider it a spectator field. In more general axion models it would be interesting to explore two-field ``spintessence''-like inflation \cite{Boyle:2001du}.} Such models are excluded at high confidence level by {\em Planck} constraints on $r$ and the scalar tilt, $n_s$. To work around this we introduce a non-minimal coupling, $\xi$, between the $s$ field and gravity (see Refs.~\cite{Okada1005,Linde1101,Kallosh1306,Kallosh1307,Linde1402,Joergensen1403,Inagaki1408} for other treatments of this model and embeddings of it in supergravity/string theory),
\begin{equation}
S_J =\int d^4x \sqrt{-g} \left[ -\left( \frac{M_{pl}^2+ \xi s^2}{2}\right) R  +\frac{1}{2}(\partial s)^2 - V(s) \right].
\end{equation}
When this action is transformed from the Jordan frame to the Einstein frame (which has a canonical gravity sector) we must define a new scalar field in order to have canonical kinetic terms, i.e. ,
\begin{equation}
S_E =\int d^4x \sqrt{-g_E} \left[ -\frac{1}{2}M_{pl}^2 R_E  +\frac{1}{2}(\partial_E \sigma)^2 - V_E(\sigma(s)) \right],
\end{equation}
where we have,
\begin{equation}
\sigma' \equiv \left( \frac{d \sigma}{d s}\right)=\frac{\sqrt{1+(s^2 \xi /M_{pl}^2) (1+6 \xi)}}{1+\ s^2 \xi/M_{pl}^2},
\end{equation}
and,
\begin{equation}
V_E=\frac{V}{(1+\xi s^2/M_{pl}^2)^2}=\frac{\frac{1}{4}\lambda \left( s^2 -f_a^2 \right)^2}{(1+\xi s^2/M_{pl}^2)^2}.
\label{eqn: potential}
\end{equation}
The slow roll parameters are then modified slightly to,
\begin{align}
\epsilon &=  \frac{1}{2} M_{pl}^2 \left(\frac{V_E'}{V_E \sigma'}\right)^2,\\
\eta &= M_{pl}^2 \left( \frac{V_E''}{V_E \sigma'^2} - \frac{V_E' \sigma''}{V_E \sigma'^3}\right).
\end{align}
\begin{figure}[tb]
\centering
\includegraphics[width=1\linewidth]{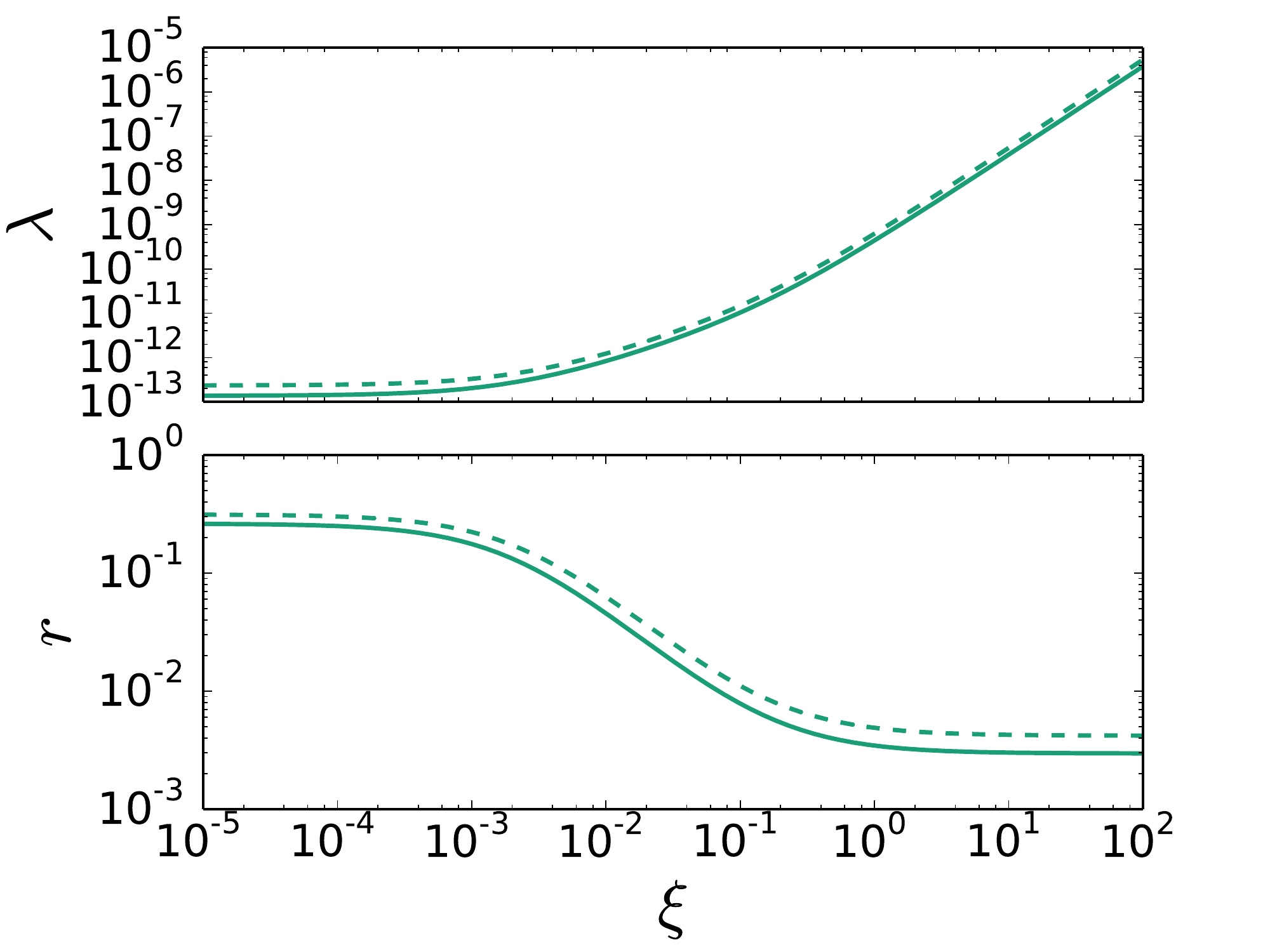}
\caption{\it The dependence of the self-coupling, $\lambda$, and the tensor-to-scalar ratio, $r$, on the non-minimal couping to gravity, $\xi$, for $N=60$ (solid) and $N=50$ (dashed). Here $\lambda$ is fixed using $A_s=2.196 \times 10^{-9}$ and we take the limit $f_a \rightarrow 0$.}
\label{fig: l and r vs xi}
\end{figure}

For the analysis of inflation in this model it is sufficient to take the limit $f_a \rightarrow 0$ in equation (\ref{eqn: potential}). For $f_a < M_{pl}$ the effect of non-zero $f_a$ is a negligible and we may treat $r$ and $f_a$ as independent (see \cite{Linde1101} for case where $f_a \rightarrow \infty$ and the effect becomes substantial). We thus have a two-parameter model of inflation. Holding the normalisation $A_s=2.196 \times 10^{-9}$ fixed reduces this to a one-parameter family of models. This is demonstrated in Fig.~\ref{fig: l and r vs xi} (upper panel) where we show the dependence of $\lambda$ on $\xi$. The values of $r$ and $n_s$ also depend on $\xi$. We show $r(\xi)$ in one dimension in Fig.~\ref{fig: l and r vs xi} (lower panel): as $\xi\rightarrow 0$, $r$ asymptotes to its value in $\lambda \phi^4$ inflation. In the opposite regime of large $\xi$ the tensor-to-scalar ratio goes to a minimum value $r_{\rm min}=3\times 10^{-3}$ for $N=60$, where $N$ is the number of $e$-folds of observable inflation.

\begin{figure}[tb]
\centering
\includegraphics[width=1\linewidth]{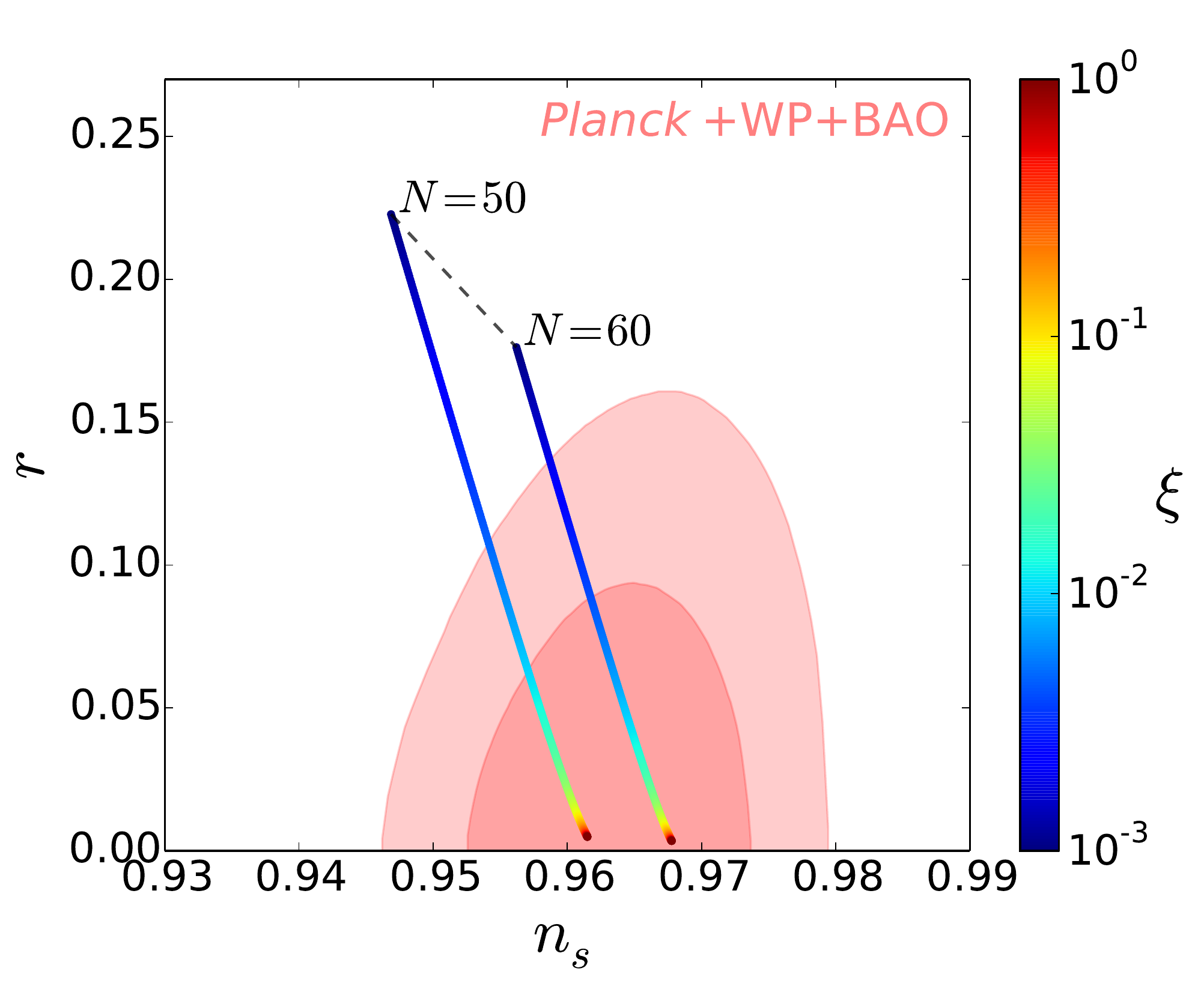}
\caption{\it The variation of the model prediction in the $n_s$-$r$ plane for different values of $\xi$. We show the 1 and 2$\sigma$ constraints from {\it Planck} with WMAP  \cite{bennett/etal:2012} polarisation (WP) and BAO from various surveys (see Ref.~\cite{PlanckCosmo} for details). Our model is consistent with the data for $\xi\gtrsim \mathcal{O}(\text{few})\times 10^{-3}$ depending on $N$, the number of $e$-folds of observable inflation.}\label{fig: ns r plot}
\end{figure}

The resulting $n_s - r$ plane predictions are shown in Fig.~\ref{fig: ns r plot}, along with the 1 and $2\sigma$ contours from {\em Planck}. Our model is flexible enough to accommodate a large part of the interesting $n_s - r$ parameter space as we await future measurements.

Apart from introducing an additional parameter, what has been the role of the non-minimal coupling? The theory without the minimal coupling resembles $\lambda \phi^4$ theory at large $s$. The potential is too steep and cannot give rise to primordial power spectra consistent with {\em Planck}. The non-minimal coupling causes the effective potential for the canonically normalised, Einstein frame field, $\sigma$, to flatten at large values of $s$ \cite{Linde1101}, allowing for large-field inflation with $r$ as a variable parameter. As the non-minimal coupling, $\xi$, is varied the model is tuned between regular quartic inflation and a copy of non-minimal Higgs inflation scenario \cite{Bezrukov0710}. With $\xi$ as a free parameter a wide range of values for $r$ can be accommodated while $s$ undergoes super-Planckian evolution and dilutes isocurvature perturbations, as we discuss below. There are observable consequences of this scenario combining axion direct detection with CMB polarisation measurements, which we will also discuss.

We note here that it is possible for quantum corrections to change the predictions of the theory. The case where the PQ scalar was also coupled to a fermion was considered in \cite{Okada1005} where the effect of this correction on the inflationary parameters was analysed. In the interest of remaining as general as possible we do not consider any such couplings. There will also be quantum corrections from the running of $\lambda$ and $\xi$ on their own (see \cite{Inagaki1408}). In our case we do not expect these corrections to have a large effect on our results because the bare coupling, $\lambda$, is very small.

\section{Constraints from the Axion Sector}
\label{sec:axion_constraints}

The cosmological evolution of the axion field is determined by the epoch in which the PQ symmetry is broken. In the standard scenario (when the radial field, $s$, plays no part in inflation) the symmetry is broken during inflation and remains broken after inflation if
\begin{equation}
f_a>\text{Max}[H/2\pi,T_{\rm rh}] \, ,
\label{eqn:iso_ineq}
\end{equation}
where $T_{\rm rh}$ is the reheat temperature. When the inequality Eq.~(\ref{eqn:iso_ineq}) is satisfied all relics of the PQ phase transition are diluted away by inflation, and isocurvature perturbations in the axion field are present. In the opposite regime
\begin{equation}
f_a<\text{Min}[H/2\pi,T_{\rm rh}] \, ,
\label{eqn:strings_ineq}
\end{equation}
then the PQ symmetry is unbroken during inflation, and no isocurvature modes are produced. However, relics of the phase transition, such as strings, now play an important cosmological role.

In our model where $s$ plays the role of the inflaton the axion acquires isocurvature perturbations regardless of the value of $f_a$ because the symmetry is always broken during inflation. The reheat temperature is then the only relevant scale in deciding whether these modes survive, and whether relics of the phase transition are cosmologically relevant.

\subsection{Isocurvature}

We begin by discussing isocurvature perturbations in the standard scenario where $s$ plays no role in inflation. 

Axions are essentially massless at energy scales $\gg \Lambda_{\rm QCD}$ so receive large quantum fluctuations [$\mathcal{O} (H/2\pi)$ every e-fold] during inflation. These fluctuations do not alter the local energy density but instead are fluctuations in the number density of axions. Axions also couple so weakly to Standard Model particles that they never return to thermal equilibrium with the rest of the Universe. As the Universe cools to below $\Lambda_{\rm QCD}$ the axion mass becomes significant and these density fluctuations must be compensated by radiation fluctuations. The fraction of axion-type isocurvature perturbations is constrained to be \cite{PlanckInflation},
\begin{equation}
\alpha= \frac{\langle (\delta T/T)_{iso}^2 \rangle}{\langle (\delta T/T)_{tot}^2 \rangle}  \lesssim 0.039,
\label{eqn: alpha bound}
\end{equation}
at $k=0.05$ Mpc$^{-1}$. 

We will use Eq.~(\ref{eqn: alpha bound}) as our constraint at arbitrary $r$, but technically this is incomplete. As we will shortly see, isocurvature constraints usually force $r\approx 0$ for axions. Axion isocurvature as constrained in Ref.~\cite{PlanckInflation} therefore assumes $r=0$, and consistent with this takes the isocurvcature spectrum to be scale invariant. Constraints on $\alpha$ will also in general be correlated with those on $r$. The combined effect of $r$ and $\alpha$ constraints on axions is so strong, however, that even an $\mathcal{O}(1)$ change to the value of either is relatively unimportant. Therefore, despite the complications just discussed, the percent-level bound of Eq.~(\ref{eqn: alpha bound}) will provide a good sense for the constraints on axion parameter space in our model including $r$.  

Axion isocurvature perturbations have (e.g. \cite{Beltran0606})\footnote{We assume that the axion accounts for 100\% of the DM relic density. This assumption is easy to drop \cite{Marsh1403}.},
\begin{equation}
\alpha=\frac{1}{1+\frac{\pi f_a^2 \langle\theta_i^2 \rangle }{M_{pl}^2 \epsilon} },
\label{eqn: alpha}
\end{equation}
where $\langle\theta_i^2 \rangle$ is fixed by the DM relic abundance for a given $f_a$. For the QCD axion we have
\begin{equation}
\Omega^{\text{mis}}_a h^2= 0.1199\left( \frac{\langle \theta_i^2 \rangle}{6 \times 10^{-6}} \right) \left( \frac{f_a}{10^{16}\ \text{GeV}}\right)^{7/6},
\label{eqn: relic density}
\end{equation}
where the angle brackets denote spatial averaging and for simplicity we have dropped anharmonic contributions to the potential and possible dilution by entropy production after the QCD phase transition (see e.g. \cite{Wantz:2009it} for more details and discussions of the accuracy and limitations of this formula). When Eq.~(\ref{eqn:iso_ineq}) is satisfied, $\langle \theta_i^2 \rangle$ must lie in the range $[\sim(H/2\pi f_a)^2,\pi^2/3]$, where the lower limit is due to back reaction of the perturbations \cite{Lyth:1991ub}. 

Using $r=16\epsilon$ and combining Eqs.~(\ref{eqn: alpha}) and (\ref{eqn: relic density}) with the {\em Planck} constraints of Eq.~(\ref{eqn: alpha bound}) and $\Omega_{\text{cdm}} h^2=0.1199 \pm 0.0027$ \cite{PlanckCosmo} we find the bound,
\begin{equation}
r \lesssim 2 \times 10^{-10} \left( \frac{f_a}{10^{16}\ \text{GeV}} \right)^{5/6}.
\label{eqn:r_bound_standard}
\end{equation}
This well-known result highlights that any conceivable detection of $r$ will put severe constraints on axion DM with GUT scale $f_a$ in the traditional setup.

If, however, the radial part, $s$, of the PQ field evolves considerably from inflation to the present, for example if it is the inflaton as we propose, the conclusion Eq.~(\ref{eqn:r_bound_standard}) can be radically changed. This is because the effective $f_{a,{\rm eff}}=s_*$ during inflation can be much larger than the vacuum value, $f_a$, appearing in the potential Eq.~(\ref{eqn:PQ_potential}). In this scenario Eq.~(\ref{eqn: alpha}) becomes
\begin{equation}
\alpha=\frac{1}{1+\frac{\pi s_*^2 \langle\theta_i^2 \rangle}{M_{pl}^2\epsilon}}.
\label{eqn:alpha_modified}
\end{equation}
In this case the $f_a$ dependence of isocurvature modes changes significantly. We see in Eq.~(\ref{eqn: alpha}) that it is usually preferable to have large $f_a$ to avoid isocurvature bounds. In our model however $f_a$ no longer directly enters the equation for $\alpha$, and it is preferable to have a smaller $f_a$ as a result of its indirect effect through $\langle\theta_i^2 \rangle$ when fixing the DM relic abundance in Eq.~(\ref{eqn: relic density}). The $r$ dependence also changes because now the important $r$-dependent quantity is $s_*^2/\epsilon$ rather than just $\epsilon$. The consequences of Eq.~(\ref{eqn:alpha_modified}) in the parameter space $(r,f_a)$ are discussed in Section~\ref{sec:conclusions}.

Another realisation of our general scheme could be achieved in volume modulus inflation \cite{Conlon:2008cj}. In string theory the axion decay constant is inversely proportional to the volume of the compact dimensions, and so if the volume evolves from small values during inflation to large values (in string units) after inflation then this too will reduce the axion isocurvature amplitude. This is achieved in Ref.~\cite{Conlon:2008cj} by inflection point inflation along the decompactification direction at small volume, with reheating occuring in a large volume meta-stable de Sitter vacuum. An attractor solution prevents the field from overshooting the meta-stable end-point. 

Since the decay constants of all axion-like particles in string theory depend inversely on the volume, the volume modulus model could dilute the isocurvature perturbations of many axions at once. In a field theory model like ours this could be achieved by inflation along a diagonal in field space with many $s$ fields, i.e. a radial-field version of N-flation \cite{Dimopoulos:2005ac}.

\subsection{Reheating}

As already noted, in our model the axion acquires isocurvature perturbations for any value of $f_a$. However, it is still the case that if the Universe reheats to a sufficiently large temperature after inflation then the PQ symmetry will be restored, eradicating the isocurvature modes. 

The PQ symmetry is restored by reheating when the thermal effective mass of the PQ field is large enough to result in an overall positive mass squared. This requires
\begin{equation}
m_{\text{eff}}^2=\frac{\lambda T_{rh}^2}{2} > \lambda f_a^2,
\end{equation}
or
\begin{equation}
T_{rh} >\sqrt{2} f_a,
\label{eqn: reheat bound}
\end{equation}
where the factor of $1/2$ is a 1-loop coefficient in the high temperature limit. 

The precise value of $T_{\text{rh}}$ is model dependent because it is determined by the coupling of the PQ field to the Standard Model (and possibly other) fields. In order to keep our discussion as general as possible we parametrise the uncertainty in $T_{\text{rh}}$ using an efficiency parameter, $\epsilon_{\text{eff}}<1$, with
\begin{equation}
T_{\rm rh}=\sqrt{\epsilon_{\text{eff}} H M_{pl}}.
\end{equation}
The phenomenology of different scenarios can then be investigated by varying $\epsilon_{\text{eff}}$. 

In the case where equation (\ref{eqn: reheat bound}) is satisfied and PQ symmetry is restored after reheating the cosmic strings that are formed when it breaks again cannot be inflated away. The decay of these cosmic strings can then produce axions and contribute to the relic density \cite{Battye9808,Yamaguchi9811,Hagmann0012,Beltran0606} with
\begin{equation}
\Omega_a^{\text{str}} h^2 \simeq (0.1 -1.0)\  7.3 \times 10^{4}\left(\frac{f_a}{10^{16}\ \text{GeV}} \right)^{1.18}\, ,
\end{equation}
where the prefactor reflects various theoretical uncertainties regarding string decay and the QCD phase transition (see \cite{Wantz:2009it} for more details). This introduces a conservative upper bound on $f_a$ in order not to over produce DM of
\begin{equation}
f_a < 1.25 \times 10^{11} \text{ GeV}\, .
\label{eqn:fa_bound_strings}
\end{equation}

\subsection{Direct Detection and Other Constraints}
The direct search for axion-like particles in the laboratory by the Axion Dark Matter Experiment (ADMX) has provided additional constraints on the parameter space. Axion DM particles with masses in the the range $m_a= (1.9-3.3)\ \mu$eV have been excluded \cite{ADMX}. We can convert this to a constraint on $f_a$ using,
\begin{equation}
m_a= \frac{\sqrt{z}}{1+z} \frac{f_{\pi} m_{\pi}}{f_a}=6.2\ \mu \text{eV} \left(\frac{10^{12}\ \text{GeV}}{f_a} \right),
\end{equation}
where $z=m_u/m_d \simeq 0.56$. This yields and exclusion in the range $f_a=(1.88-3.26)\times 10^{12}$ GeV.

There also exists an upper bound on the axion mass of $m_a \lesssim 10^3\ \mu$eV ($f_a \gtrsim 6.2 \times 10^9$ GeV) due to astrophysical limits on the axion-photon coupling \cite{Raffelt0611}. Larger axion masses result in a large photo-axion coupling and can significantly alter the cooling time of stars, radiation from SN1987A, solar neutrino flux and other phenomena. A lower bound on the mass of the QCD axion follows from the phenomenon of black hole super radiance and the observed spins of stellar mass blackholes, excluding $f_a\gtrsim 10^{17}\text{ GeV}$ \cite{axiverse2009}.

\section{Results and Conclusions}
\label{sec:conclusions}
\begin{figure}
\centering
\includegraphics[width=1\linewidth]{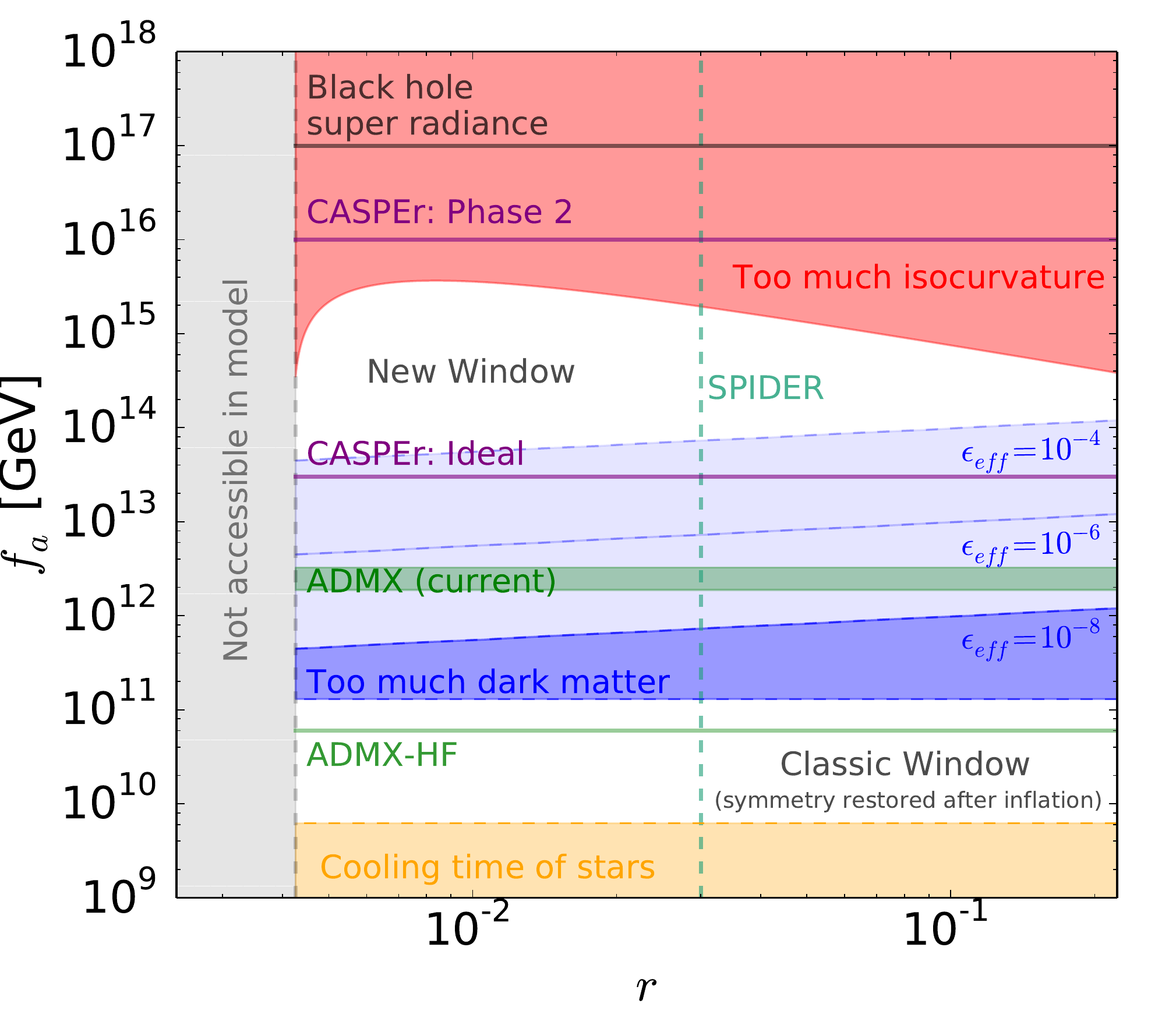}
\caption{\it Axion DM constraints for non-minimal PQ inflation model showing the new window unavailable to other axion models. The red region is ruled out by isocurvature constraints. The orange region is rule out by astrophysical constraints on the photon-axion coupling. The blue region is ruled out by overproduction of DM from cosmic strings (shown for three different reheating temperatures parameterised by $\epsilon_{\text{eff}}$). The green region is excluded by direct searches for DM axions by ADMX. The purple lines show the projected lower bounds of the CASPEr experiment. Together, \textsc{Spider} and CASPEr/ADMX-HF can probe a large part of the parameter space of our model.}\label{fig: fa r constraints}
\end{figure}
The results of this paper are summarised in Fig.~\ref{fig: fa r constraints} where we show the constraints on $f_a$ as a function of the tensor-to-scalar ratio in our model of inflation driven by the radial PQ field. The upper portion of the plot is ruled out by excess isocurvature modes for any observable value of $r$ even when our mechanism is employed. Our mechanism opens up a new window for intermediate-scale axions with $10^{12}\text{ GeV}\lesssim f_a\lesssim 10^{15}\text{ GeV}$ to be consistent with observable primordial $B$-modes, as could be observed, for example, by near future experiments like \textsc{Spider}. 

The exact size of the new window depends on the value of $r$, which has a minimum value, $r_{\rm min}\approx 4\times 10^{-3}$ in our model. This is below what is accessible to \textsc{Spider}, but it is not impossible to imagine this as detectable at some stage in the future. In standard inflation $r\lesssim 10^{-10}$ is required for high $f_a$ axions to be viable in the so-called anthropic window \cite{Hertzberg0807}. If $r$ were detected, for example by 21cm lensing, in the range $10^{-9}\lesssim r\lesssim 10^{-3}$ then a mechanism other than ours would be necessary to save the high $f_a$ QCD axion.

Remaining agnostic about the model of reheating and allowing $\epsilon_{\rm eff}$ to vary by orders of magnitude has a strong effect on the size of the new window, with the lower bound of the window $f_{a,{\rm low}}\propto \epsilon_{\rm eff}^{0.5}$. Even when reheating is quite efficient (up to $\epsilon_{\text{eff}} \sim 10^{-2}$) our model is still able to accommodate large values of $r$ and $f_a$ simultaneously within some window. The size of the new window is maximised when reheating is inefficient and the blue region disappears; this occurs for $\epsilon_{\text{eff}} \lesssim 10^{-10}$. 

We have also highlighted the presence of the classic window for axion DM, when the PQ symmetry is restored after inflation. Here the lower bound on $f_a$ is imposed by astrophysical constraints \cite{Raffelt0611}, while the upper bound is imposed by the DM abundance from string decay. When the reheating is very inefficient ($\epsilon_{\text{eff}} \lesssim 10^{-10}$) the size of the classic window can be reduced significantly because the symmetry cannot be restored.

The ADMX exclusion lies in the new window so we can look forward to more explorations of this window (and the classic window) with the proposed ADMX-HF experiment \cite{vanBibber1304} that will extend the sensitivity to masses as large as $\sim 100\ \mu$eV ($f_a \sim 6\times 10^{10}$ GeV). The CASPEr experiment \cite{CASPEr} has proposed a search for axions with large $f_a$ using the precession of $CP$-odd nuclear moments of target sample caused by interacting with DM axions. Phase 2 of the experiment can rule out axions with $f_a > 1.3 \times 10^{16}$ GeV. With improvements in magnetometer technology the experiment can be used to search for axions with $f_a > 4 \times 10^{13}$ GeV. Without some mechanism to dilute isocurvature, such as ours, the entire range for CASPEr would be excluded on cosmological grounds if $r$ is observed by \textsc{Spider}.

We have shown that it is possible for large $f_a$ axion DM to coexist with high scale inflation, with observably large tensor modes and accompanying isocurature. If a non-negligible measurement of $r$ is reported in future by e.g. Keck-Array \cite{KeckArray} or \textsc{Spider} this would be selective in the available parameter space of our model. Furthermore if large $f_a$ axions are found by CASPEr or ADMX then a mechanism such as that presented in this paper will be needed to reconcile the two measurements. Additional probes of the model could come if isocurvature perturbations are observed at the percent level by future CMB polarisation measurements  \cite{hamann2009}. Axion DM direct detection and CMB polarisation experiments are complementary in many ways and together can access physics at extremely high energies and discriminate between models of inflation.
\section*{Acknowledgements}
We are grateful to Joe Conlon for useful discussions, and to University of Oxford for hospitality during these discussions. MF is grateful for funding provided by the UK Science and Technology Facilities Council. RH is supported by the KCL NMS graduate school. DJEM's research at Perimeter Institute is supported by the Government of Canada through Industry Canada and by the Province of Ontario through the Ministry of Economic Development \& Innovation.  
\bibliographystyle{h-physrev}
\bibliography{bib_axion}
\end{document}